\documentclass[superscriptaddress,twocolumn,showpacs,preprintnumbers,amsmath,amssymb,letterpaper,prX]{revtex4-1}

\usepackage{graphicx}
\usepackage{dcolumn}
\usepackage{bm}
\usepackage{color}
\usepackage{ulem}

\newcommand{\re}[1]{\text{Re}[#1]}
\newcommand{\be}{\begin{equation}}
\newcommand{\ee}{\end{equation}}

\begin{document}

\title{Pump-Controlled Modal Interactions in Microdisk Lasers}

\author{Seng Fatt Liew}
\affiliation{Department of Applied Physics, Yale University, New Haven, CT 06520, USA}
\author{Li Ge}
\affiliation{Department of Engineering Science and Physics, College of Staten Island, CUNY, Staten Island, NY 10314, USA}
\affiliation{The Graduate Center, CUNY, New York, NY 10016, USA}
\author{Brandon Redding}
\affiliation{Department of Applied Physics, Yale University, New Haven, CT 06520, USA}
\author{Glenn S. Solomon}
\affiliation{Joint Quantum Institute, NIST and University of Maryland, Gaithersburg, MD 20899, USA}
\author{Hui Cao}
\email{hui.cao@yale.edu}
\affiliation{Department of Applied Physics, Yale University, New Haven, CT 06520, USA}
\date{\today}

\begin{abstract}
We demonstrate an effective control of nonlinear interactions of lasing modes in a semiconductor microdisk cavity by shaping the pump profile.
A target mode is selected at the expense of its competing modes either by increasing their lasing thresholds or suppressing their power slopes above the lasing threshold.
Despite of strong spatial overlap of the lasing modes at the disk boundary, adaptive pumping enables an efficient selection of any lasing mode to be the dominant one, leading to a switch of lasing frequency.
The theoretical analysis illustrates both linear and nonlinear effects of selective pumping, and quantify their contributions to lasing mode selection.
This work shows that adaptive pumping not only provides a powerful tool of controlling the nonlinear process in multimode lasers, but also enables the tuning of lasing characteristic after the lasers have been fabricated.
\end{abstract}


\maketitle

\section{Introduction}
Multimode lasers are a classical example of nonlinear open systems that display complex behaviors. Recently there have been growing interests and efforts in understanding and manipulating the properties and interactions of lasing modes \cite{harayama_PRL03,conti_PRL06, Science, nicolas_1,lopez_NatComm13, nicolas_2, ge_NP}. For example, non-Hermitian physics of exceptional points have been explored for the control of lasing in coupled cavities \cite{EP,EPexp1,EPexp2,EP_CMT}. The Parity-Time symmetry, or more specifically, the balanced gain and loss distribution in space, has been utilized for lasing mode selection \cite{ring0,ring1,ring2}. These processes, however, only target for modifying the linear properties of individual lasing modes, including their respective thresholds and power. It would be fascinating, but more difficult, to manipulate their nonlinear interactions. Lately active control of pump profile for random lasers has been proposed and demonstrated using the spatial light modulator (SLM) \cite{lopez_NP,nicolas_1,nicolas_2,lopez},
and it was found that non-uniform pumping enables a highly complex nonlinear modal interaction.

While these studies show it is possible to realize a preferably single-mode, frequency switchable, on-chip coherent light sources by controlling the pump profile, several fundamental questions must be addressed first, including (1) whether each of the lasing modes can be modified by the pump profile similar to that in random lasers; (2) if lasing modes are strongly overlapped in space, whether a desired mode can be selected via the pump control while the others are suppressed; and (3) how important the contributions from the linear effect (spatial overlap between the pump and the mode) and from the nonlinear effect (mode competition and gain saturation) are for lasing mode selection.

In this work we present a thorough experimental and theoretical study in the platform of the semiconductor microdisk laser, which has been utilized for single-photon emitters and biochemical sensors in integrated photonic circuits \cite{Microcavity1,Microcavity2} and shown potential tunability as a pump-controlled device \cite{liew}. Such feature is unexpected from the results of previous studies, which rely on the enhanced spatial overlap of the target mode with the pumping pattern or the pumped-induced modification of the spatial mode profile for lasing mode selection.


First, while the modes of a weakly scattering random laser can adapt themselves to the spatially inhomogeneous pumping pattern to facilitate lasing in the target mode, the lasing modes in a semiconductor microdisk are much more rigid, as they are tightly confined by total internal reflection at the disk boundary. Typically the lasing modes correspond to the whispering-gallery (WG) modes with a low radial number, thanks to their high quality ($Q$) factor and the resulting low lasing threshold. These modes are barely modified by the pump profile, as confirmed experimentally from their emission patterns \cite{liew}. Therefore, pump-controlled tunability cannot be achieved through the modification of each individual mode in a microdisk laser.

Second, the only possibility remaining for realizing a pump-controlled microdisk laser is then through the differentiation of individual lasing modes. Lasing in a selected mode may be facilitated by reducing its lasing threshold via increasing its spatial overlap with the pump profile \cite{pereira,chenJOB01,chen,bisson,naidoo,rex,chern,harayama,chern2,narimanov,hentschel1,hentschel2}. To differentiate multiple lasing modes then requires little spatial overlap of this desired mode with all other modes, otherwise lasing in non-selected modes would be enhanced as well. Unfortunately, the high-$Q$ WG modes in a microdisk have very strong spatial overlap at the disk boundary, especially the ones with the same radial number, making mode selection almost impossible unless subwavelength control of the pump profile can be realized. This pessimistic outlook, however, apply only to a perfectly round microdisk. In practice, there is always inherent cavity surface roughness introduced unintentionally during the fabrication process. The coupling between high-$Q$ WG modes of low radial numbers and lower-$Q$ WG modes of higher radial numbers, caused by the surface roughness \cite{ge_PRA13,ge_PRA1,redding_PRL14}, enhances the spatial diversity of the lasing WG modes, which plays a critical role to differentiate the WG modes by adaptive pumping.

Experimentally we find that selective pumping changes not only the lasing threshold but also the power slope of each WG modes in a semiconductor microdisk, which shows that both the linear and nonlinear effects caused by the spatially inhomogeneous pump profile are important. To suppress lasing in all non-selected modes, the threshold of the selected mode often increases too. The selected mode is not necessarily the first mode to lase (i.e. with the lowest lasing threshold), instead it may grow the fastest and become the dominant lasing mode at high pumping level.
Our theoretical analysis is able to differentiate the linear and nonlinear effects of selective pumping, and quantify their contributions to lasing mode selection.

Below we discuss these findings in detail, demonstrating that the adaptive pumping not only provides a powerful tool
for controlling the nonlinear process in multimode lasers, but also enables the tuning of lasing characteristic after the lasers have been fabricated. We first present the experimental findings and then analyze them using the semiclassical laser theory.

\section{Adaptive pumping of a microdisk laser}

\begin{figure}[htbp]
	\centering
	\includegraphics[width=\linewidth]{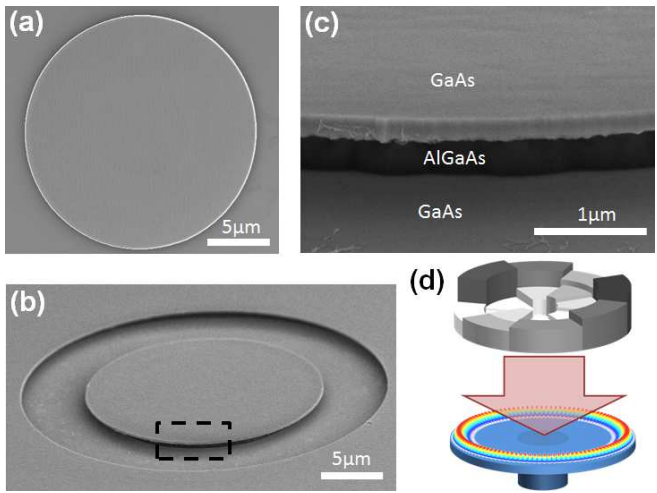}
	\caption{Fabricated GaAs microdisk. (a) Top-view and (b) tilt-view scanning electron microscope images of a GaAs microdisk. (c) Magnified view of the part highlighted in (b) reveals the sidewall roughness of the disk. (d) A schematic showing the spatial intensity profile of the pump beam is modulated to control the lasing modes in the microdisk. }
	\label{fig1}
\end{figure}

The experiments were conducted on circular microdisks of larger radius and smaller boundary deformation than the ones used in Ref.~\cite{liew}. The larger radius reduces the free spectral range, and hence increases the number of possible lasing modes within the same gain bandwidth. Although many more modes managed to lase under uniform pumping, we were able to make every single one of them the dominant lasing mode with adaptive pumping. Interestingly we observed different scenarios for the selected modes to win the competition, as described below.

 A 200-nm-thick GaAs layer and a 1000-nm thick Al$_{0.75}$Ga$_{0.25}$As layer were grown on a GaAs substrate by molecular beam epitaxy.
Three layers of InAs quantum dots were embedded in the middle of the GaAs layer.
The circular disks of radius 10 $\mu$m were patterned by electron-beam lithography and followed by two steps of etching.
The first is a inductively-coupled-plasma reactive etching with a BCl$_3$ and Cl$_2$ mixture to create GaAs/Al$_{0.75}$Ga$_{0.25}$As cylinders. The second is a selective HF-based etching of the Al$_{0.75}$Ga$_{0.25}$As to create a pedestal underneath the disk.
Figure \ref{fig1} shows the scanning electron microscope (SEM) images of a fabricated disk.
From the top-view SEM image [Fig. 1(a)],  the disk shape has negligible deformation from a circle, and the disk radius is 9.2 $\mu m$.
The high-magnification tilt-view SEM image [Fig. 1(c)] reveals the roughness of the disk sidewall.
As will be discussed later, such disorder is crucial to the realization of lasing mode control that is otherwise impossible for a perfectly smooth disk.

\begin{figure}[htbp]
	\centering
	\includegraphics[width=\linewidth]{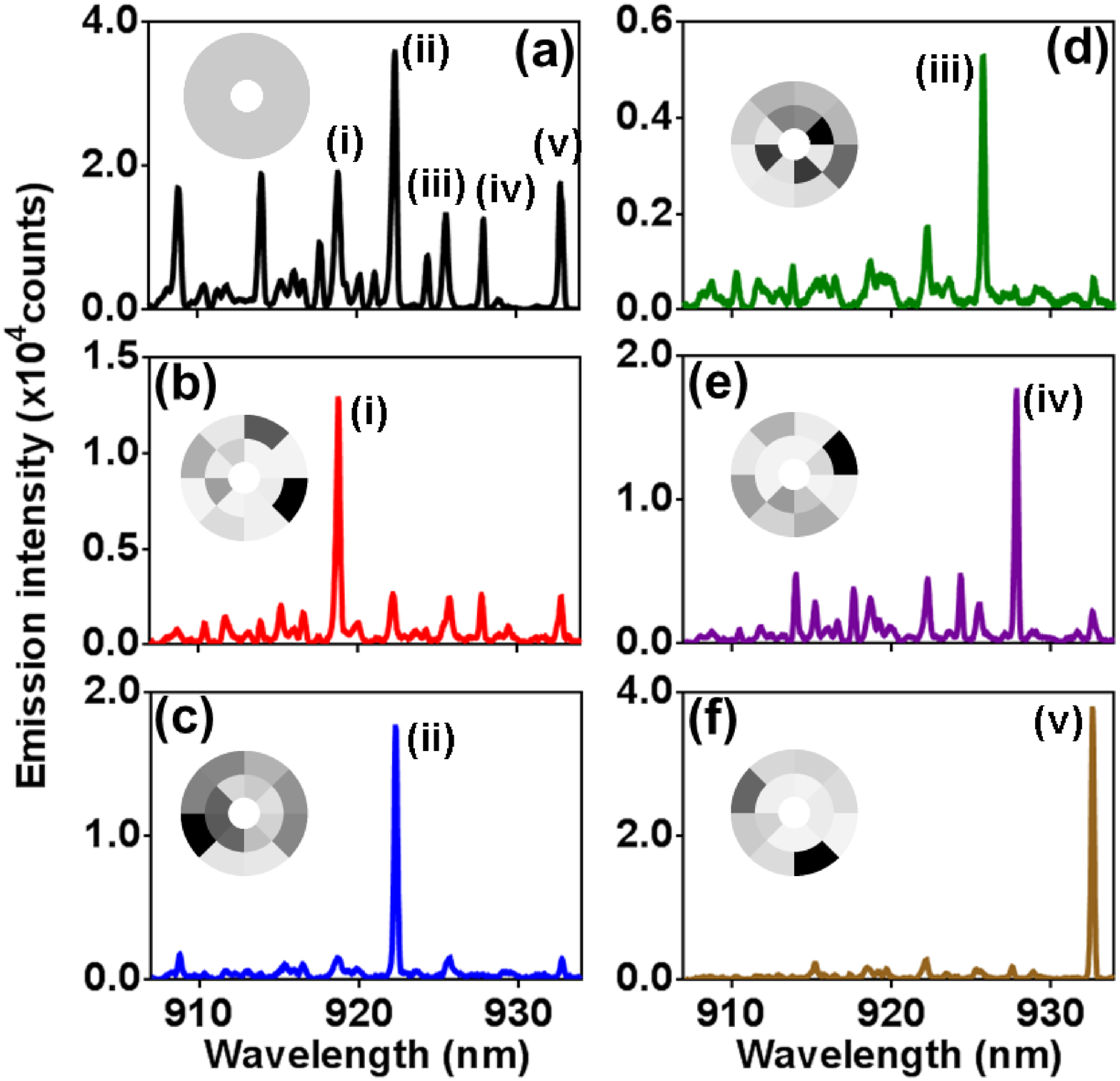}
	\caption{Lasing spectra with different pump patterns. (a) Under uniform ring pumping, many modes lase simultaneously. (b-f) Lasing spectra when modes (i-v) are target as the dominant lasing mode with adaptive pumping. Insets show the final optimized pump profiles in gray scales with darker color corresponding to higher intensity.}
	\label{fig2}
\end{figure}

The lasing experiment was conducted on individual microdisk mounted in a low-temperature cryostat.
The experimental setup was similar to the one described in ref. \cite{liew}.
The InAs QDs were optically excited by a mode-locked Ti:Sapphire laser ($\lambda_p$ = 790nm, 76MHz, 200fs pulses).
As shown schematically in Fig. 1(d), the spatial intensity distribution of the pump beam was modulated by a spatial light modulator (Hamamatsu X10468-02), and then projected onto the top surface of a microdisk by a long-working distance objective lens.
The emitted light was scattered out of the disk plane at the disk edge, some of it was collected by the same objective lens and focused into a multimode fiber which was connected to a spectrometer.
Since the high-$Q$ modes avoid the central region of the disk where light could leak to the substrate via the pedestal, we set the pump region to a ring as shown in the inset of Fig. 2(a).
When the pump light was uniformly distributed across the ring, many modes lased simultaneously, as seen in the main panel of Fig. \ref{fig2}(a).

As mentioned before, any naive attempt to select one WG modes by shaping the spatial pump profile will unavoidably enhance other WG modes with similar spatial  profiles. Therefore, to make any one of the lasing modes to be the dominant, we also need to suppress all others. To achieve this, we adopt the Genetic Algorithm and shape the pump profile iteratively \cite{liew}.
The cost function chosen for optimization is the extinction ratio $G = I_m/I_o$, where $I_m$ is the peak intensity of the target mode and $I_o$ is the highest intensity among all other modes.
This cost function leads to optimization of the pump profile for the highest contrast between the target mode intensity and the competing mode intensities.
The annular pump region on the disk is divided into two subrings, each is further divided into eight sections in the azimuthal direction.
Such coarse meshing precludes any spatial selectivity of ideal WG modes in a perfectly circular disk that have the same radial number but different azimuthal number, as their azimuthal modulations of intensity differ only on a much finer length scale.
The lasing modes, at least some of them, are expected to be WG modes with the lowest radial number, because they have the highest $Q$ factor and lowest lasing threshold.
Thus it would have seemed impossible to separate these modes by enhancing the spatial overlap of the pump pattern with only one of them.

Strikingly, the adaptive pumping scheme enabled any of the lasing modes to dominate over the others.
Figure 2(b-f) show the examples of the lasing modes labeled (i-v) in Fig. 2(a).
Each of the panels is the lasing spectrum obtained after optimization for one of the modes, and the inset is the optimized pump profile in gray scales with darker colors representing higher intensity.
The total pump power is kept at a constant value of 4 mW during the optimization process, thus the pump power is only re-distributed to different region of the disk.
After about 300 iterations of optimization,  one of the modes (i-v) has become the dominant lasing modes with extinction ratio $G > 2.8$.

We repeated the experiment using different cost functions that either enhances the target mode intensity $I_m$ without suppressing others  or reduces the non-selected mode intensity $I_o$ while ignoring the change of $I_m$.
In both cases, all lasing modes are enhanced or suppressed simultaneously, without mode selection.
For example, the optimization of $I_m$ focused the pump light around the disk boundary where all the lasing modes are concentrated.
This results illustrate that the choice of an appropriate cost function is crucial to the realization of mode selection.

\begin{figure}[b]
	\centering
	\includegraphics[width=\linewidth]{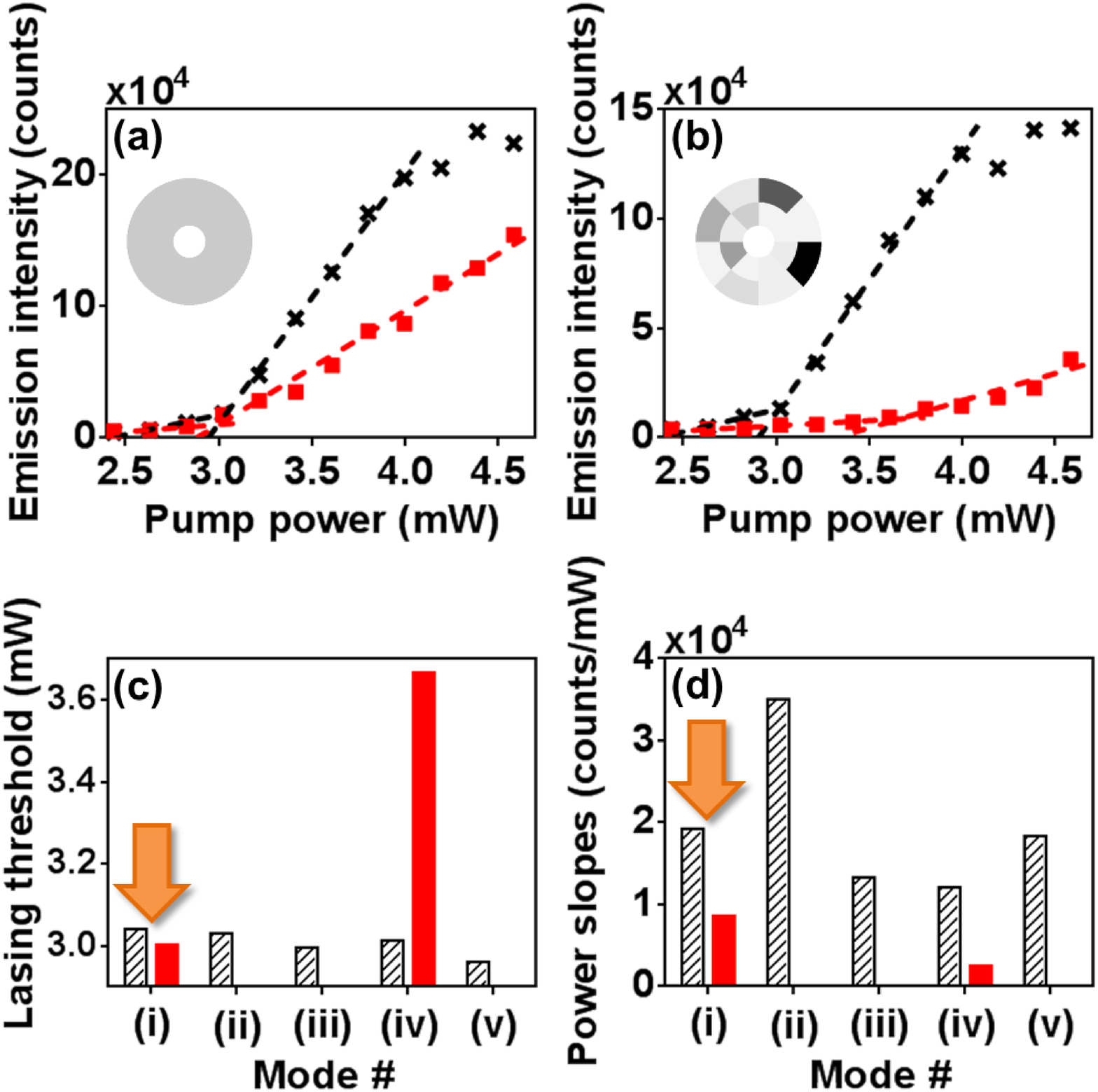}
	\caption{Emission intensity as a function of pump power for (a) mode (i) and (b) mode (iv) under uniform ring pumping (black crosses) and non-uniform pumping (red squares) to optimize mode (i). The insets show the intensity distribution of (a) uniform and (b) optimized pump profiles. Dashed lines represent the linear regression of the data points below and above the lasing threshold. The intersection of the two lines determines the lasing threshold pump power, and the slope of the line above the threshold gives the power slope of the lasing mode.  (c,d) Lasing thresholds and power slopes for modes (i-v) under uniform pumping (patterned bars) and adaptive pumping (filled bars). Mode (i) becomes the dominant lasing mode because the inhomogeneous pumping suppress all other competing modes by greatly increasing their lasing thresholds. }
	\label{fig3}
\end{figure}

To understand the mechanism of mode selection, we measured the emission intensity of individual mode as a function of the pump power for the optimized pump profile and compared to the uniform pumping.
Figure \ref{fig3} shows the data for the selection of mode (i).
Under the uniform ring pumping, mode (i) displays a sharp increase in the growth rate of its emission intensity with the pump power $P$ at $\sim 3$ mW, indicating the onset of lasing action.
When $P$ exceeds 4 mW, the growth rate is reduced due to gain depletion.
With the optimized pump profile, mode (i) turns on at almost the same pump power, but it grows at a significantly reduced rate than the uniform pumping.
Thus the adaptive pumping does not enhance lasing in the selected mode, instead it suppresses lasing in the non-selected modes.
As an example, Fig. 3(b) plots the emission intensity of mode (iv) in the two pumping cases.
Under the uniform pumping it behaves similarly to mode (i), namely, it turns on at the pump power of 3 mW and becomes saturated beyond 4~mW. With the optimized pump profile, however, its emission is greatly reduced, as it turns on at a higher pump power of 3.65~mW and grows at a much lower rate.
The other lasing modes, e.g. modes (ii), (iii) and (v), experience more suppression and do not turn on even when the pump power is increased to 5 mW.

\begin{figure}[b]
	\centering
	\includegraphics[width=\linewidth]{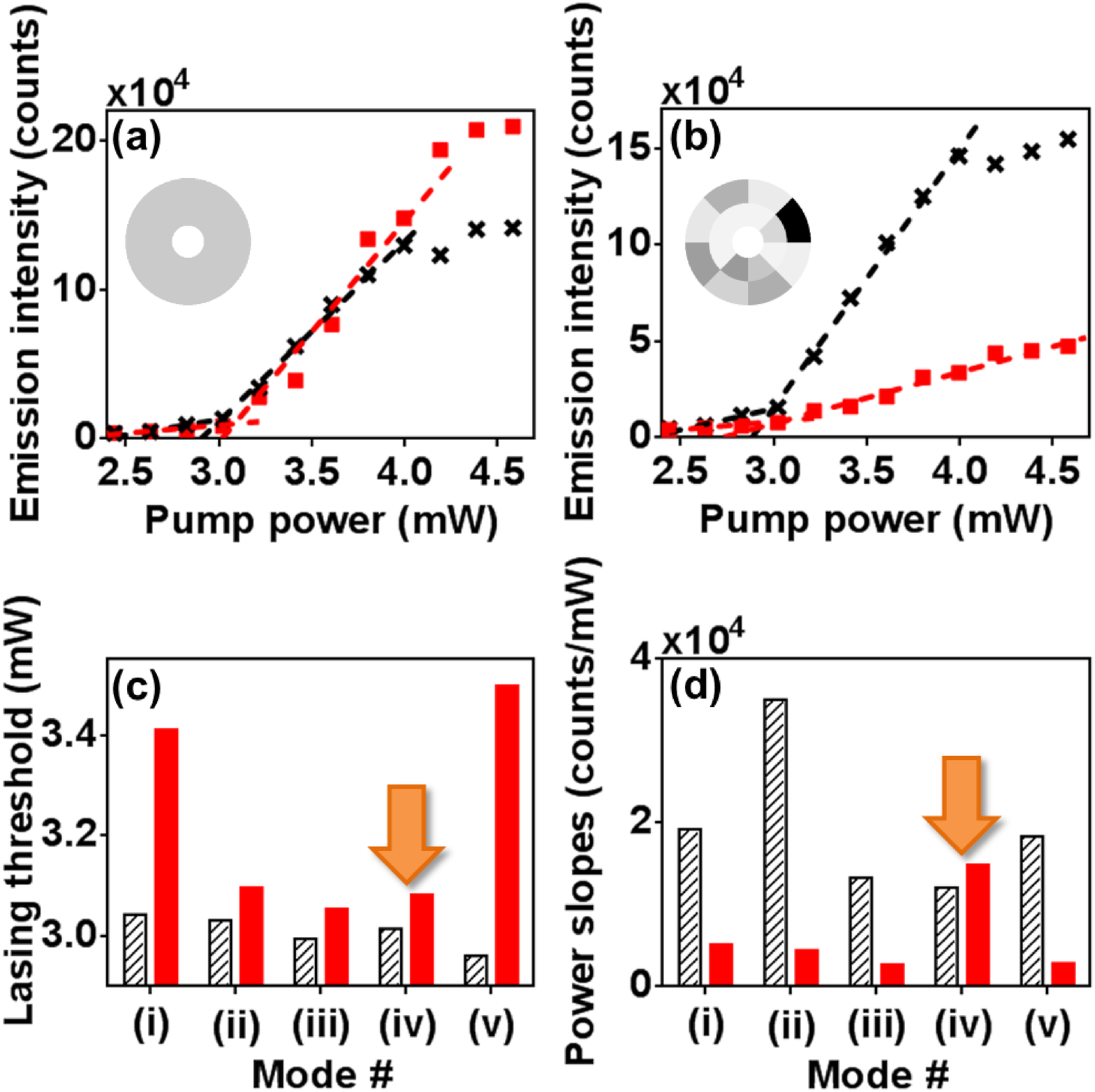}
	\caption{Emission intensity as a function of pump power for mode (iv) in panel (a) and mode (iii) in panel (b) under uniform ring pumping (black crosses) and non-uniform pumping to select mode (iv) (red squares). The insets show the intensity distribution of (a) uniform and (b) optimized pump profiles. (c,d) Lasing thresholds and power slopes under uniform pumping (patterned bars) or selective pumping (filled bars). The target mode (iv) becomes the dominant lasing mode even though its lasing threshold is not the lowest because adaptive pumping greatly suppress the power slopes of the competing modes . }
	\label{fig4}
\end{figure}

To be more quantitative, we extract the lasing threshold and power slope of individual modes from the data.
The linear fitting of the emission intensity with the pump power before and after a mode turns on gives the lasing threshold (the pump power at which the two linear lines intersect) and the power slope (the gradient of the linear fit above the threshold).
The lasing thresholds and power slopes of modes (i-v) are plotted in Fig. \ref{fig3}(c) and (d) for uniform and selective pumping respectively.
All five modes (i-v) have similar lasing threshold under uniform pumping, but their power slopes exhibit more variation.
After optimizing the pump profile to select mode (i), the lasing thresholds of all non-selected modes increase significantly.
Even though its threshold is reduced only slightly by the adaptive pumping, the selected mode (i) has the lowest lasing threshold and becomes the first one to turn on.
These results suggest the optimized pump profile has reduced spatial overlap with the non-selected modes, while keeping the spatial overlap with the selected mode almost the same.
Given that all the high-$Q$ WG modes should overlap strongly at the disk boundary, the spatial discrimination achieved by the optimized pump profile is remarkable and unexpected, which will be addressed in the next section.

Increasing the difference in lasing threshold is not the only scenario to achieve mode selection, in some cases the target mode does not have the lowest lasing threshold, but the highest power slope instead.
This is seen when mode (iv) is chosen to be the dominant lasing mode (Fig. \ref{fig4}).
With the optimized pump profile, mode (iv) is not the first one to lase, but it grows the fastest with increasing pump power above the lasing threshold, and its intensity exceeds all other modes at the pump power where the optimization is conducted.
For example, mode (iii) has the lowest lasing threshold under selective pumping, so it lases first, but its power slope is much less than that with uniform pumping.
In fact, the power slopes of all non-selected modes are reduced significantly by adaptive pumping, and become much smaller than that of the selected mode.
In contrast, the power slope of the selected mode increases slightly and the gain depletion appears at a higher emission intensity.
These results suggest that the selective pumping modifies nonlinear interaction of the lasing modes and favors the selected mode in their competition for gain.

\begin{figure}[t]
	\centering
	\includegraphics[width=\linewidth]{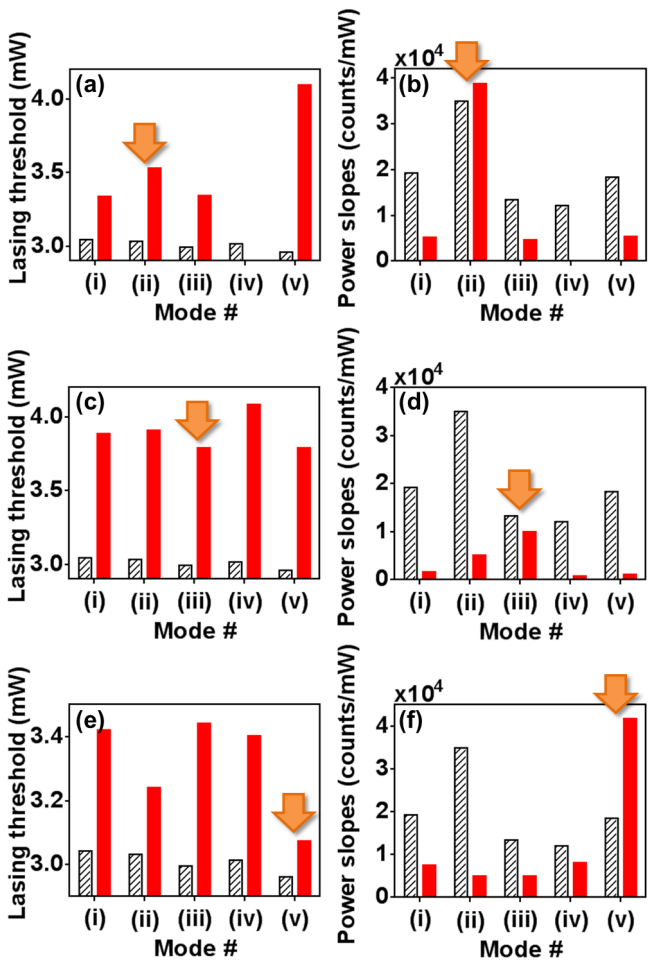}
	\caption{(a,c,e) Lasing thresholds and (b,d,f) power slopes for modes (i-v) after optimizing for mode (ii), (iii) and (v). Patterned bars represent the case for uniform pumping and filled bars for selective pumping. While the target mode not necessarily has the lowest lasing threshold, their power slope becomes the highest among all lasing modes. }
	\label{fig5}
\end{figure}

Figure \ref{fig5} shows the results for optimization of modes (ii), (iii) and (v).
The selection of mode (ii) leads to a similar behavior to mode (iv): the target mode has the highest power slope instead of the lowest lasing threshold, and the power slopes of all competing modes are reduced dramatically [Fig. \ref{fig5}(a,b)].
In contrast, the selection of mode (iii) greatly increases the lasing thresholds of all modes including the target one and reduces their power slopes.
Nevertheless, the selected mode (iii) becomes the dominant lasing mode because its power slope experiences the least reduction as compared to all other modes especially mode (v) which has a similar lasing threshold [Fig. \ref{fig5}(c,d)].
To select mode (v), the lasing thresholds of all modes increase but the increment is the smallest for the target mode and makes it the first lasing mode [Figure \ref{fig5}(e)].
While the power slopes of the competing modes (i-iv) decrease as their thresholds increase, the power slope of mode (v) increases by a factor of two in spite of the increase of its lasing threshold.
These results illustrate that the adaptive pumping is not only capable of selecting different modes to lase first by making their lasing threshold the lowest in the linear regime (up to the first lasing threshold), but can also efficiently control the multimode lasing process in the highly nonlinear regime well above threshold.

\section{Theoretical analysis of lasing mode selection}

The remarkable degree of control of lasing process in the high-$Q$ microcavity with strong mode overlap is unprecedented.
To interpret these results, we adopt a numerical model based on the Steady-state Ab-initio Laser Theory (SALT) \cite{TS,Science,ge_PRA10} to compute the lasing mode intensity under different pumping conditions.
The nonlinear modal interactions included in the original form of SALT assumes an homogeneously broadened gain medium, so that all the lasing modes interact with the same population inversion and a strong lasing mode may deplete the gain available for others.
The gain medium in our microdisk laser is inhomogeneously broadened, as the different sizes of InAs QDs lead to distinct emission frequencies.
Since the inhomogeneous width is larger than the homogeneous width, the QDs may be grouped within each homogeneous width, and a lasing mode interacts with only one group that is in resonance with its own frequency.
This seems to suggest two lasing modes with frequency spacing larger than the QD homogeneous width would interact with different groups of QDs and do not interact.
However, experimentally the pump light at $\lambda$ = 790 nm generated electron-hole pairs in the GaAs layer, which subsequently relaxed to the InAs QDs.
Hence, all the QDs shared the same reservoir of carriers.
When a lasing mode rapidly depleted the carriers in one group of QDs, these QDs would get replenished quickly from the GaAs layer, reducing the available carriers for other modes.
Thus the lasing modes, even with frequency spacing larger than the QD homogeneous width, competed for the carriers in the shared reservoir \cite{strauf_PRL}.
This competition induces nonlinear modal interactions that can be effectively captured by the SALT.

Since the high-$Q$ modes in the GaAs microdisk have low lasing thresholds, they usually lase and their frequencies and spatial profile are barely changed by the pump.
Thus each lasing mode can be approximated by a single quasi-bound mode in the passive cavity \cite{TS}; and
the interactions of lasing modes, through the gain medium, only affects their lasing thresholds and intensities above threshold.
This ``single-pole" approximation (SPA) simplifies the SALT formula to the following set of equations \cite{ge_PRA10}
\be
\frac{D_0}{D_0^{(\mu)}} - 1 = \sum_{\nu} \Gamma_{\nu}\chi_{\mu\nu} I_{\nu}, \label{eq:SPASALT}
\ee
which are linear in terms of the pump strength $D_0$ and the lasing mode intensities $I_\mu$.
$D_0^{(\mu)}$ is the threshold of mode $\mu$ for a given pump profile $f(\vec{r})$ and in the absence of modal interactions.
If $f(\vec{r})$ is normalized by $\int_p f(\vec{r}) d\vec{r} = A$, where the subscript ``$p$" denotes integration over the ring-shaped pump region of area $A$, $D_0^{(\mu)}$ is simply given by \cite{ge_NP}
\be
D_0^{(\mu)}\approx\frac{n^2}{Q_\mu \re{f_\mu}},\quad f_\mu \equiv \int_c f(\vec{r}) \Psi^2_\mu(\vec{r}) d\vec{r},\label{eq:TH}
\ee
where the subscript ``$c$" denotes integration over the entire microdisk. Here $n$ is the refractive index of the microdisk,  $Q_\mu$ and $\Psi_\mu(\vec{r})$ are the $Q$-factor and wave function of mode $\mu$, respectively. $\Psi_\mu (\vec{r})$ is normalized using $\int_c \Psi^2_\mu(\vec{r}) d\vec{r}=1$.
As evident in Eq.~(\ref{eq:TH}), the direct outcome of n on-uniform pumping is modifying the non-interacting threshold $D_0^{(\mu)}$ through the pump overlapping factor $f_\mu$.
In general $f_\mu$ is a complex number, but for the high-$Q$ modes in the microdisk it is approximately a real number.
For uniform pumping across the ring, $f(\vec{r})$ is equal to 1 inside the ring and 0 outside.  
Thus $f_\mu = 1$, since the high-$Q$ WG modes have little overlap with the central region of the disk that is on top of the pedestal.
$\Gamma_\nu$ on the right hand side of Eq.~(\ref{eq:SPASALT}) is the Lorentzian gain factor of the lasing mode $\nu$, which can be taken as 1, because each lasing mode interacts only with the resonant QDs as mentioned above.
The interaction matrix $\chi$ contains the self-interaction coefficients ($\mu = \nu$) and cross-interaction coefficients ($\mu \neq \nu$) of the lasing modes.
Because the lasing modes in the microdisk cavity correspond to the high $Q$-modes, $\chi$ can be defined as
\be
\chi_{\mu\nu} = \frac{1}{A}\left\vert \int_c d\vec{r} \, \Psi_\mu^2(\vec{r}) \, |\Psi_\nu(\vec{r})|^2 \right\vert.\label{eq:chi}
\ee
Note that the interaction matrix itself does not change with the pump profile $f(\vec{r})$; the effect of the latter is reflected by the changes in the lasing threshold and power slope above threshold.

Excellent agreement has been obtained between the results of the SPA-SALT equations (\ref{eq:SPASALT}) and the direct numerical solution to the Maxwell-Bloch equations in previous studies \cite{ge_PRA10,N_level}.
The latter approach is extremely demanding in terms of computational power and time, which is impractical for our analysis here.
We note that the SALT incorporates the modal interactions to infinite order, which is preserved in the SPA.
To illustrate how the SPA-SALT captures the modification of gain saturation by selective pumping, let us consider the simplest case of single mode lasing.
The power slope of the lasing mode (mode 1), defined by $d I_1/d D_0$, can be found from Eqs.~(\ref{eq:SPASALT}) and (\ref{eq:TH}),
$S_1 = [\chi_{11}D_0^{(1)}]^{-1} = Q_1 \re{f_1}/\chi_{11} n^2$.
Its value is proportional to the pump overlapping factor and inversely proportional to the self-interaction coefficient.
Thus changing the pump profile will modify the power slope of the lasing mode via nonlinear self-saturation of gain.

In the discussion below, we will show that the mechanisms of different mode selection scenarios observed experimentally can be illustrated with the interactions of two modes only, and we will focus on the role of non-uniform pumping.
The lasing threshold and power slope of the second lasing mode strongly depends on its interaction with the first lasing mode.
Because of this interaction, the threshold of the second mode is not $D_0^{(2)}$ but rather
\be
D_{0,int}^{(2)} = \frac{\chi_{11}/\chi_{21}-1}{\chi_{11}/\chi_{21}-D_0^{(2)}/D_0^{(1)}} D_0^{(2)}. \label{eq:D2}
\ee
We will refer to $D_{0,int}^{(\mu)}$ as the interacting threshold, and $D_0^{(\mu)}$ the non-interacting threshold.
For the first lasing threshold, $D_{0,int}^{(1)} = D_{0}^{(1)}$.
The second lasing mode has a higher non-interacting threshold, $D_0^{(2)} > D_0^{(1)}$,  and its interacting threshold $D_{0,int}^{(2)}$ is further increased by the interaction between mode 1 and 2, $D_{0,int}^{(2)} > D_{0}^{(2)}$.
The power slope of the second lasing mode is given by
\be
S_{2,int} = \frac{\chi_{11}/\chi_{21}-D_0^{(2)}/D_0^{(1)}}{\chi_{11}/\chi_{21}- \chi_{12}/\chi_{22}} S_2, \label{eq:S2}
\ee
where $S_2 = [\chi_{22}D_0^{(2)}]^{-1} = Q_2 \re{f_2}/\chi_{22} n^2$  is its power slope in the absence of interaction with mode 1.
Due to the modal interaction, the power slope of the first lasing mode is changed to
\be
S_{1,int} = \frac{\chi_{22}/\chi_{12}-D_0^{(1)}/D_0^{(2)}}{\chi_{22}/\chi_{12}- \chi_{21}/\chi_{11}} S_1, \label{eq:S1}
\ee
once the second mode starts lasing.

To see how the nonlinear modal interaction can be controlled by selective pumping, let us consider two pump profiles $f(\vec{r})$ and $\tilde{f}(\vec{r})$.

According to the above equations, different pump profiles will change both the lasing threshold and power slope of the second lasing mode through the  non-interacting threshold $D_0^{(\mu)}$.
The resulting changes of both quantities reflect the nonlinear effects induced by the modification of pump profile.
If there were no nonlinear interaction between mode 1 and 2, by modifying the pump profile from a particular $f(\vec{r})$ to another one $\tilde{f}(\vec{r})$, the lasing threshold of the second mode would change from $D_0^{(2)}=n^2/Q_2f_2$ to $\tilde{D}_0^{(2)} = n^2/Q_2\tilde{f}_2$, with $\tilde{f}_2$ defined by having $\tilde{f}(\vec{r})$ in Eq.~(\ref{eq:TH}).
However, the actual threshold of mode 2 is given by Eq.~(\ref{eq:D2}) instead, with $D_0^{(\mu)}$ replaced by the corresponding $\tilde{D}_0^{(\mu)}$, and
its dependence on the non-interacting threshold of the first lasing mode reflects the modal interaction that varies with the pump profile.
Similarly, the power slopes of both modes, given by Eqs.~(\ref{eq:S2}) and (\ref{eq:S1}), depend on the ratio of $D_0^{(1)}$ and $D_0^{(2)}$, indicating
both self-saturation and cross-saturation effects can be modified by the pump profile.

To quantify the change of the nonlinear effects due to the variation of the pump profile, we use the ratio of the power slopes $\tilde{S}_{2,int}$ [with pump profile $\tilde{f}(\vec{r})$] and $S_{2,int}$ [with pump profile $f(\vec{r})$] for mode 2:
\be
{\cal M}_2 \equiv \frac{\tilde{S}_{2,int}}{S_{2,int}}
= \frac{\tilde{f}_2\chi_{11}/\chi_{21}-Q_1\tilde{f}_1/Q_2}{f_2\chi_{11}/\chi_{21}-Q_1f_1/Q_2}. \label{eq:M2}
\ee
We note that this modification factor ${\cal M}_2$ also governs the change of the interacting threshold of mode 2, i.e. ${D_{0,int}^{(2)}}/{\tilde{D}_{0,int}^{(2)}} ={\cal M}_2$.
For mode 1, the ratio of its power slopes after mode 2 lases is
\be
{\cal M}_1 \equiv \frac{\tilde{S}_{1,int}}{S_{1,int}}  = \frac{\tilde{f}_1\chi_{22}/\chi_{12}-Q_2\tilde{f}_2/Q_1}{f_1\chi_{22}/\chi_{12}-Q_2f_2/Q_1}. \label{eq:M1}
\ee
Therefore, we see that besides the pump overlapping factors and the $Q$-factors, two important modal interaction parameters are the ratios of the interaction coefficients - $\chi_{11}/\chi_{21}$ and $\chi_{22}/\chi_{12}$.

Next we will apply the above formula to the analysis of lasing mode selection by adaptive pumping.
If mode 1 is the dominant lasing mode with uniform pumping, there are two scenarios to make mode 2 dominant by non-uniform pumping.
The first one is to have mode 2 lase first by lowering its non-interacting threshold to below that of mode 1, similar to the scenario shown in Fig. 3.
Since $D_0^{(\mu)}$ scales inversely with the pump overlapping factor, non-uniform pumping may enhance the pump overlap with mode 2 and/or reduce the pump overlap with mode 1.
This requires little spatial overlap between mode 1 and 2, which is not true for the high-$Q$ WG modes in a perfect circular microdisk.
However, the disk sidewall roughness, albeit small, has a profound effect on the mode profiles, as we will illustrate with numerical simulation.

Due to the finite resolution of the scanning electron microscope (SEM), we could not map the exact boundary roughness to compute the lasing modes observed experimentally.
Also a disk of radius $\sim$ 10 $\mu$m supports many high-$Q$ modes, and the experimental uncertainty about the disk temperature during optical pumping leads to an inaccurate estimation of the refractive index,  making the matching of lasing frequencies between simulation and experiment extremely difficult.
As a compromise, we chose to simulate a smaller disk with radius 3 $\mu m$ to provide a physical understanding of the effects of adaptive pumping.
The effective index of refraction of the disk is set to $ n = 3.13$ \cite{qinghai_PRA}.
We introduced surface roughness on the disk boundary by adding random high order harmonic perturbations,
the disk boundary is described in the polar coordinates as $\rho(\theta) = R + \sum_{m=20}^{80}a_m\cos(m\theta +\phi_m)$, where
$a_m$ and $\phi_m$ are random numbers in the range [-0.5 nm, 0.5 nm] and $[-\pi, \pi]$, respectively.
The highest order harmonic perturbation, $m = 80$, is chosen to be slightly larger than $2\pi n R/\lambda$, because light in the cavity cannot resolve boundary modulation much finer than its wavelength $\lambda/n$.
The lowest-order harmonic perturbation, $m = 20$, limits the scale of boundary modulation to be less than 1 $\mu$m, which is consistent with the SEM images of the fabricated microdisks.

\begin{figure}[t]
	\centering
	\includegraphics[width=\linewidth]{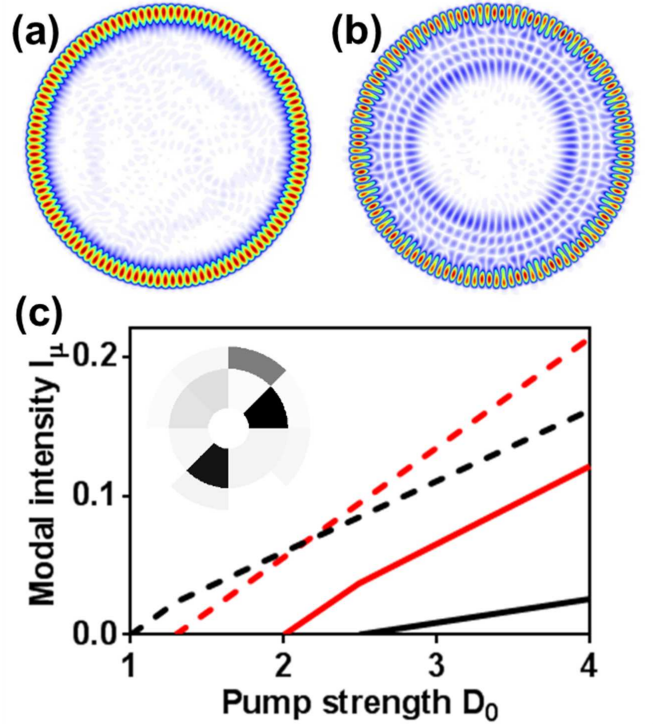}
	\caption{Numerical simulation of switching of lasing order by adaptive pumping of a microdisk with rough boundary. The disk radius is 3 $\mu$m, and the refractive index is 3.13. (a,b) Calculated spatial distribution of magnetic field ($|H_z|$) for two TE-polarized high-$Q$ modes, labeled mode 1 and 2, respectively. Mode 1, at $\lambda$ = 917.7 nm, has $Q$ factor of $1.20 \times 10^5$. Mode 2, at $\lambda$ = 902.9 nm , and $Q = 1.09 \times 10^5$.  Mode 1 in (a) consists predominantly of CCW propagating WG wave with radial number 1 and azimuthal number 57. Mode 2 in (b) is primarily a WG modes of radial number 1 and azimuthal number 58, but contains additional WG waves of higher radial order number 7 due to disorder-induced mode mixing. (c) Lasing intensity for mode 1 and 2 as a function of pump strength $D_0$ with uniform ring pumping (dashed lines) or non-uniform pumping to select mode 2 (solid lines). Mode 1 lases first before mode 2 with uniform pumping, but the lasing order switches, i.e., mode 2 lases first, with the non-uniform pump profile (inset) shown in the gray scale with darker color corresponding to stronger pumping. $D_0$ is normalized to the lasing threshold of mode 1 with uniform ring pumping.}
	\label{fig6}
\end{figure}

We calculated the quasi-bound modes of the passive cavity using the finite-element frequency-domain method \cite{COMSOL}.
Two of the high-$Q$ modes with transverse electric (TE) polarization (electric field parallel to the disk plane) are shown in Fig.~\ref{fig6} (a,b).
Their $Q$-factors are $Q_1\simeq1.20\times10^5$ and $Q_2\simeq1.09\times10^5$, respectively.
They are both WG modes, but exhibit additional features than the regular WG modes.
In mode 1 [Fig.~\ref{fig6} (a)], the counter-clockwise (CCW) propagating WG wave dominates over the clockwise (CW) wave, reducing the contrast between the field intensity maxima and minima.
By expanding the intracavity field distribution in the cylindrical harmonics, we identify its radial number is 1 and azimuthal number 57.
Similar analysis of mode 2 [Fig.~\ref{fig6} (b)] reveals that in addition to the primary WG wave with radial number 1 and azimuthal number 58, it has additional WG waves of higher radial order number 7 and azimuthal number 34.
This is attributed to the mixing of WG modes with different order via wave scattering by the rough boundary of the microdisk \cite{ge_PRA13,ge_PRA1,redding_PRL14}.

Using these mode profiles, we find that the interaction parameters are $\chi_{11}/\chi_{21}$ = 1.64 and $\chi_{22}/\chi_{12}$ = 1.44.
Under uniform ring pumping, $f_1=f_2 = 1$, and $D_0^{(2)}/D_0^{(1)} = Q_1/Q_2$ = 1.1.
Mode 1 lases first due to its higher $Q$-factor as illustrated in Fig. \ref{fig6}(c), where the pump strength $D_0$ is normalized to the first lasing threshold $D_0^{(1)}$.
We numerically simulated adaptive pumping using the same polar grid as in experiment (double rings, each divided further into 8 regions of equal area).
For each pump pattern, we first calculated the pump overlapping factor and the non-interacting threshold for the two modes, then solved Eq.~(\ref{eq:SPASALT}) to find the modal intensities.
To enhance lasing in mode 2, we ran the Genetic algorithm to maximize the intensity ratio of mode 2 over mode 1, $I_2/I_1$, by redistributing the pump energy across the ring while keeping the total pump strength fixed at $D_0 = 3$.
The adaptive pumping switches the order of lasing, as shown in Fig.~\ref{fig6} (c), and mode 2 lases first.
The pump overlapping factor for mode 1 is reduced to $\tilde{f}_1 = 0.47$, and mode 2 to $\tilde{f}_2 = 0.55$.
Both modes have higher lasing thresholds due to reduced spatial overlap with the pump.
Nevertheless, the optimized pump pattern has larger overlap with mode 2 by shifting most pump energy to the inner ring.
This is only possible because mode 2 extends more into the inner ring by disorder-induced mixing of different WG modes.
The larger pump overlapping factor of mode 2 overcomes its lower $Q$ factor, $\tilde{f}_2/\tilde{f}_1 > Q_1/Q_2$, making $D_0^{(2)} < D_0^{(1)}$.

\begin{figure}[b]
	\centering
	\includegraphics[width=\linewidth]{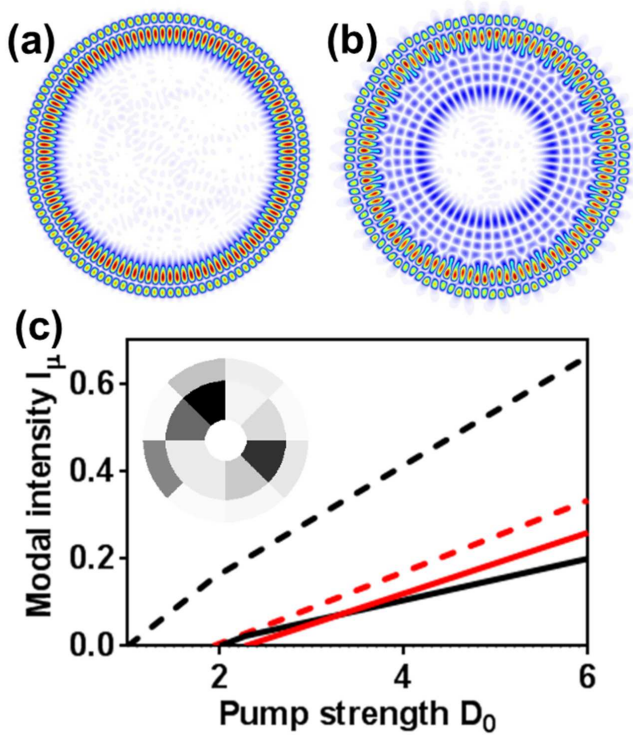}
	\caption{Numerical simulation of changing the power slopes of lasing modes by adaptive pumping of a microdisk with rough boundary. The disk radius is 3 $\mu$m, and the refractive index is 3.13. (a,b) Calculated spatial distribution of magnetic field ($|H_z|$) for two WG modes of second radial order, labeled mode 1 and 2, respectively. (c) Lasing intensity for these two modes with uniform ring pumping (dashed lines) and non-uniform pumping to make mode 2 stronger than mode 1 (solid lines). The optimized pump profile found by the Genetic Algorithm does not change the lasing order but greatly suppress the power slope of mode 1 (black lines).}\label{fig7}
\end{figure}

The second scenario of making the second lasing mode stronger than the first is to modify their power slopes without switching the order of lasing.
This is similar to the scenario observed in Fig.~\ref{fig4}, and the selectivity was enabled by the strong suppression of the power slopes of all other lasing modes, including the one with the lowest threshold (mode 1 here).
In Fig.~\ref{fig7}(a,b) we consider another pair of high-$Q$ modes with TE polarization, and their $Q$-factors are $Q_1\simeq1.31\times10^5$ and $Q_2\simeq1.05\times10^5$, respectively.
By decomposing their intracavity field distributions by cylindrical harmonics, we confirm that both modes consists primarily of a second radial order WG modes.
In addition, mode 2 contains additional field components from another nearby low-$Q$ WG modes of a higher radial order, due to disorder-induced mode mixing.
Using their mode profiles, we find that the interaction parameters are $\chi_{11}/\chi_{21}$ = 1.71 and $\chi_{22}/\chi_{12}$ = 1.43.

Under uniform ring pumping ($f_1 = f_2 = 1$), $D_0^{(2)}/D_0^{(1)}$ is given by $Q_1/Q_2$ = 1.25, and mode 1 turns on first due to its higher $Q$-factor.
Its power slope is higher than that of mode 2, as shown in Fig. \ref{fig7}(c) with the pump strength $D_0$ normalized by $D_0^{(1)}$.
After the optimization process to maximize $I_2/I_1$ at $D_0 = 3$, the pump overlap factors are changed to  $\tilde{f}_1 = 0.5$ and $\tilde{f}_2 = 0.59$.
Even though the optimized pump profile has a stronger overlap with mode 2 by concentrating pump energy to the inner ring [inset of Fig. \ref{fig7}(c)], the change is not sufficient to compensate the difference in the $Q$ factor, i.e. $\tilde{f}_2/\tilde{f}_1 < Q_1/Q_2$, and mode 1 remains the first lasing mode as $\tilde{D}_0^{(1)} < \tilde{D}_0^{(2)}$.
Nevertheless, the power slope of mode 2, calculated with Eq.~(\ref{eq:S2}), is reduced only by $1-{{\cal M}_2}=17\%$ according to Eq.~(\ref{eq:M2}).
In contrast, the power slope of mode 1 is reduced by $1-{{\cal M}_1}=61\%$, which is much more than mode 2.
Consequently, the power slope of mode 2 becomes higher than mode 1 with non-uniform pumping, thus its lasing intensity exceeds that of mode 1 at $D_0 > 3.3$.
This result shows that adaptive pumping can enhance lasing in the target mode through nonlinear modal competition above threshold and modify the power slopes of lasing modes significantly.
Such capability could not be achieved by the conventional pump engineering of semiconductor microcavity lasers \cite{pereira,chenJOB01,chen,bisson,naidoo,rex,chern,harayama,chern2,narimanov,hentschel1,hentschel2}.

\section{Discussion and Conclusion}

For a smooth microdisk without boundary roughness, we have verified numerically that adaptive pumping does not work for the corresponding WG modes in the previous examples or any other high-$Q$ WG modes with same radial order.
These modes have very similar spatial profiles which cannot be differentiated by the coarse modulation of the pump region, and hence preclude any spatial selectivity.
For example, the Genetic algorithm could not find an optimized pump profile that switches the lasing order of the two WG modes (without disorder) in Fig. \ref{fig6} because the ratio of their pump overlapping factors $\tilde{f}_2/\tilde{f}_1$ remains the same for different pump patterns.
Compared to a smooth microdisk, the disorder on the disk boundary results in the generation of diversified mode profiles with different pump overlapping factors and interaction coefficients, as confirmed in our numerical simulation.
Such diversity is expected to scale up with the disk size, as the density of high-$Q$ WG modes increases, and the reduced frequency spacing facilitates the mixing of WG modes with different radial order.
Thus the larger disks used in our experiment have more versatile interactions among the lasing modes, which can be controlled by shaping the spatial pump profile.

Compared to the weakly scattering random lasers which have a large spatial degree of freedom, the high-$Q$ microcavity lasers inherently have less degree of control.
Nevertheless, we demonstrate that the intrinsic fabrication imperfection enables the diversified modal interactions that can be utilized to control the lasing behavior.
We believe our method can be extended to electrical pumping by patterning the contact electrodes and controlling the current injection, creating a tunable on-chip light source for integrated photonic applications.

We thank Patrick Sebbah, Nicolas Bachelard, Stefan Rotter, Douglas Stone, Alex Cerjan, and Jan Wiersig for stimulating discussion.
This work is supported by the MURI grant No. N00014-13-1-0649 from the US Office of Naval Research and by the National Science Foundation under the Grant No. DMR-1205307.

\end{document}